# Precision Cutting and Patterning of Graphene with Helium Ions

D.C. Bell[1,2], M.C. Lemme[3], L. A. Stern[4], J.R. Williams[1,3], C. M. Marcus[3]

1. School of Engineering and Applied Sciences, Harvard University, Cambridge MA 02138
2. Center for Nanoscale Systems, Harvard University, Cambridge MA 02138
3. Department of Physics, Harvard University, Cambridge MA 02138
4. Carl Zeiss SMT Inc., Peabody MA 01960

**Abstract**

We report nanoscale patterning of graphene using a helium ion microscope configured for lithography. Helium ion lithography is a direct-write lithography process, comparable to conventional focused ion beam patterning, with no resist or other material contacting the sample surface. In the present application, graphene samples on Si/SiO$_2$ substrates are cut using helium ions, with computer controlled alignment, patterning, and exposure. Once suitable beam doses are determined, sharp edge profiles and clean etching are obtained, with little evident damage or doping to the sample. This technique provides fast, lithography compatible with graphene, with ~ 15 nm feature sizes.

## Introduction

Helium ion microscopy is a recently developed high-resolution imaging technology useful for a variety of materials applications, such as materials that would normally charge under an electron beam and especially imaging of carbon nano-structures [1]. Being a charged ion beam instrument, it is also possible to perform milling and sputtering, as commonly done with a gallium focused ion beam (FIB) system. Advantages of helium ion lithography (HIL) compared to FIB include its ability to mill and sputter soft and fragile materials at low rates and its small probe size (< 0.5 nm) [2].

Graphene is a two-dimensional hexagonal lattice of carbon atoms with thickness of one or a few atomic layers [3]. Due to its material stability and strength, absence of defects, and unique electronic band-structure, graphene holds considerable promise for a number of applications in nanoscale electronics, optoelectronics, and mechanics, as well as being of fundamental interest in condensed matter physics. Many potential applications, such as high-speed field-effect transistors, require graphene to be patterned at the nanoscale. Unlike comparable structures made from carbon nanotubes, patterned graphene can form complex extended geometries and can be readily contacted electrically, yielding a well-controlled connection between micron-scale and nanometer-scale systems and devices.

Existing methods of patterning graphene include electron beam lithography in conjunction with reactive ion etching (e.g. [4][5][6]) and direct etching with a FIB in a transmission electron microscope (TEM) [7]. Both methods are suitable to produce patterns in the tens of nanometer range. The former is limited by uncontrolled under-etching by the oxygen plasma; the latter relies on transferring graphene flakes onto TEM grids, which is not suitable for larger scale fabrication of devices. The work reported here focuses on process considerations of HIL applied to graphene lying on a substrate. Elsewhere, application of HIL to graphene-based electronic devices, including *in situ* electrical measurement during lithography on graphene including suspended graphene, is presented [8].

**Process Considerations**

Helium ion microscopy is based on field ionization of helium ions using a cryogenically cooled tungsten tip, which is truncated by a trimer of atoms (Fig. 1). The gun is centered so that ion emission from a single atom of the trimer is used for imaging. The beam current can be modified by changing the imaging gas pressure, with typical operation in the range of fA to pA. Details of operation have been described elsewhere [2].

The interaction of the ion beam with the sample determines the resolution of the instrument as a microscope as well as the minimum size of an etched line when used as a lithography tool. Figure 2. illustrates the interactions of primary energetic He ions with a graphene layer on $SiO_2$ substrate indicating the production of sputtered ions, secondary electrons ($SE_I$ at the primary beam and $SE_{II}$ at the secondary scattered ion exiting the surface), back scattered ions (BSI) and secondary ions (SI).

The size of the interaction region within the sample depends on the atomic mass and density of the sample, and on the acceleration voltage. The interaction area at the sample surface is smaller for HIL than for commercial electron-beam lithography or FIB systems by roughly a factor of 100.

Figure 3 shows a comparison of gallium and helium ion beam propagation into graphene on a $Si/SiO_2$ substrate, calculated using the TRIM software package [10][11]. For graphene on a typical substrate of 285 nm $SiO_2$ on silicon, the simulation shows the gallium atoms depositing most of their kinetic energy in the upper most parts of the material (Fig. 3(a)). While this makes gallium highly effective at bulk milling and etching, the resulting surface interaction area gives feature sizes much larger than the actual beam diameter. TRIM calculations for helium ions on an identical specimen, in contrast, show that 99.6% of the ions pass directly through the graphene with no ion interaction at all. Instead, the majority of the ion energy is deposited deep within the silicon substrate (Fig. 3(b)). These simulations suggest that the lighter mass and higher speed of the helium ions results in smaller interaction volume with the surface layers and hence better resolution and potential milling feature size. From the perspective of sputtering and patterning, the result is a reduced proximity effect in the surface layer. The light ion mass results in low energy transfer and hence a relatively lower sputtering yield compared to gallium.

Figure 3(c) shows a simulation for a suspended layer of graphene over $SiO_2$ and silicon substrate. The thickness of the vacuum gap and the $SiO_2$ add up to 285 nm, as would be the case in a device fabricated from a typical substrate as in Fig. 3b. In this case, there is little to no observable backscattering to the graphene layer. Experimental details of our suspended graphene etching by helium ions have been described elsewhere [8] the inclusion of the modeling Fig. 3(c) of suspended graphene is to show, for completeness the inherent differences of each etching approach. In addition, the lack of interaction of backscattered ions with the graphene film should make suspended devices particularly suitable for He ion etching.

Overall, these simulations suggest a factor of 10 reduction in feature size for helium ions compared with gallium ions. Gallium ions also could leave ionic contamination in samples which are problematic for graphene devices, Long-term trapping of helium ions is expected to be much less severe. An important result from these simulations is the indication that the lack of helium ion beam divergence in the vicinity of the surface of the sample down to a depth of about 100 nm should enable nanometer scale fine etching and cutting, with minimal surrounding area damage. Following from the simulations [11], a model of proximity effect during ion bombardment showing the differences between gallium and helium ion bombardment for milling and etching (Insets in Fig. 3).

**Experiment**

Helium ion lithography and microscopy was carried out using a Zeiss ORION system, operated at 30kV acceleration voltage with a beam current of 1 to 1.6 pA. While hydrocarbons have been used previously to write patterns onto graphene [12], here such contamination is avoided: The chamber is cleaned with air plasma overnight prior to sample patterning using an Evactron plasma cleaner at 12 W with a least 10 on/off (15 minutes/45 minutes) cycles [13]. Beam control for HIL used a modified Nabity Nanometer Pattern Generation System (NPGS). The Nabity system allowed for dose computer-based pattern design, dosage variations, and alignment to existing features, including sample edges [8]. Test writing was performed on 285 nm-thick $SiO_2$ on a Si substrate. Initial dose exposures indicated a dose of 1.2 nC/cm as an optimal initial setting for ion beam and dwell times in the pattern generation system. Atomic force microscope (AFM) and helium ion microscopy images (Fig. 4(a,b)) shown sharp, well-defined patterned etched in $SiO_2$. AFM data was taken in tapping mode using a Digital Instruments Nanoscope III instrument. We used Veeco Tapping Mode Etched Silicon Probes (TESP) with a specified nominal tip radius of curvature of 5-10 nm

Graphene flakes were then deposited onto the $SiO_2$ by mechanical exfoliation, similar to the method described in [3] with the modifications of the process as described in [8]. Next, mono- and multi-layer graphene flakes were identified with an optical microscope.

**Results and Discussion**

In an initial experiment, a ion beam spot was focused on a $SiO_2$ supported multi-layer graphene flake, resulting in small holes in the material. Fig. 5 shows a helium ion micrograph of one such hole with a diameter of ~ 15 nm. Variations of this dose were performed to ascertain the optimal operation point for He ion etching. Figure 6(a) shows a helium ion micrograph of lines etched in graphene sample, showing changes in the pattern with increasing beam dose from left to right at a measured probe current of 1.6 pA. The dose was varied from 3 nC/cm to 15 nC/cm in 3 nC/cm steps. Total patterning time varying between 3 and 6 seconds. The result indicates that a suitable dose for etching a graphene sample with the present set-up is 10-15 nC/cm. A larger dose variation performed with 20, 100 and 200 nC/cm is shown in Fig. 6(b). The scanning electron microscope (SEM) image shows that all doses lead to a cut in the single-layer graphene. However, the combination of SEM and AFM images further reveals that for very high doses the underlying substrate can swell by at least 50 nm from the effect of ion knock-on damage to the underlying silicon. The detailed AFM analysis for the single-layer graphene cut with the low dose rate of 20 nC/cm for test lines resulted in a measured depth of 4 nm (Fig. 7).

NPGS software allows patterning of highly complex structures, as an example this is demonstrated in Fig. 8 with a Harvard logo etched into a multi-layer graphene flake with line widths of order 15 nm. The overall dimensions of the logo are about 4 μm x 5 μm.

**Conclusions**

We have shown that it is possible to precisely cut and pattern graphene using 30kV helium ions using a modified helium ion microscope, configured with a commercial beam-patterning package that allows control of beam dose, pattern configuration, and alignment to existing features. We have demonstrated the technique by patterning single-layer and multilayer graphene samples, yielding sub-20nm feature sizes. We expect that this technique will facilitate graphene-based electronic devices that take advantage of the unique physical properties of graphene.

**Acknowledgements**

MCL acknowledges the support of the Alexander von Humboldt foundation through a Feodor Lynen Research Fellowship. Research also supported in part by the NRI INDEX program. The authors thank S. Nakaharai for fruitful discussions regarding the process.

**References**

[1] D.C. Bell, "Contrast Mechanisms and Image Formation in Helium Ion Microscopy", Microscopy and Microanalysis, 15 (2), pp 147-153, 2009.

[2] L. Scipioni, L.A. Stern, J. Notte, S. Sijranddij, B.J. Griffin, "Helium ion microscope". Adv Mater Proc 166, 27–30. 2008.

[3] K. S. Novoselov, A. K. Geim, S. V. Morozov, D. Jiang, Y. Zhang, S. V. Dubonos, I. V. Grigorieva, and A. A. Firsov, "Electric Field Effect in Atomically Thin Carbon Films", Science, 306, pp. 666-669, October 2004.

[4] M.Y. Han, B. Özylimaz, Y. Zhang, P. Kim, "Energy band-gap engineering of graphene nanoribbons", Phys. Rev. Lett., Vol. 98, 2007.

[5] Z. Chen, Y.-M. Lin, M. J. Rooks, P. Avouris, "Graphene nano-ribbon electronics", Physica E, Vol. 40, Iss. 2, 2007.

[6] L.A. Ponomarenko, F. Schedin, M.I. Katsnelson, R. Yang, E.W. Hill, K.S. Novoselov, A.K. Geim, "Chaotic Dirac Billard in Graphene Quantum Dots", Science, Vol. 320, 2008.

[7] M.D. Fischbein, M. Drndić, "Electron beam nanosculpting of suspended graphene sheets", Appl. Phys. Lett. 93, 113107, 2008.

[8] M.C. Lemme, D.C. Bell, J.R. Williams, L.A. Stern, B.W.H. Baugher, P. Jarillo-Herrero, C.M. Marcus, "Etching of Graphene Devices with a Helium Ion Beam", DOI arXiv:0905.4409 May 2009.

[9] J.R. Williams, L. DiCarlo, C.M. Marcus, "Quantum Hall Effect in a Gate-Controlled p-n Junction of Graphene", Science, 317(5838):638-641, 2007.

[10] R.S. Averbeck, M. Ghaky, "A model for Surface Damage in Ion-Irradiated Solids", Journal of Applied Physics, 76 6, 3908 (1994).

[11] J. Zeigler, J. Biersack, U. Littmark, "The Stopping Range of Ions in Matter. New York Pergamon Press (also at http://srim.org)

[12] J.C. Meyer, C.O. Girit, M.F. Crommie, A. Zettl, "Hydrocarbon lithography on graphene membranes", Appl. Phys. Lett. 92, 123110, 2008.

[13] R Vane, "Immobilization and Removal of Hydrocarbon Contamination Using the Evactron De-Contaminator", Microscopy and Microanalysis, **12**,(S02), pp. 1662-1663, 2006.

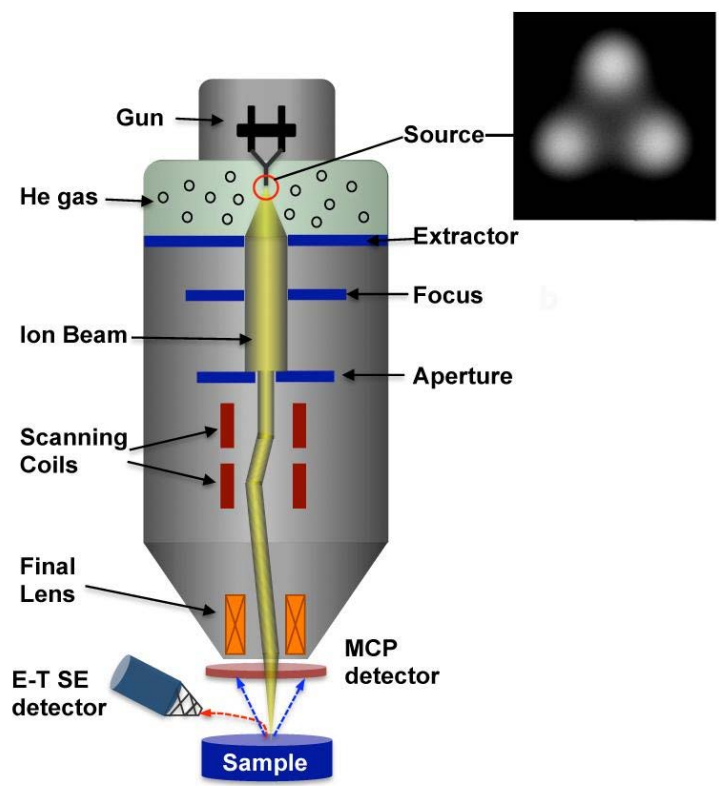

**Figure 1.** Schematic of the Zeiss ORION helium ion microscope column, showing the ion source, apertures, Everhart-Thornley secondary electron (E-T SE) detector and microchannel plate (MCP) detector configuration.  (Inset) Image of atomic trimer on a tungsten tip, where ionization occurs.

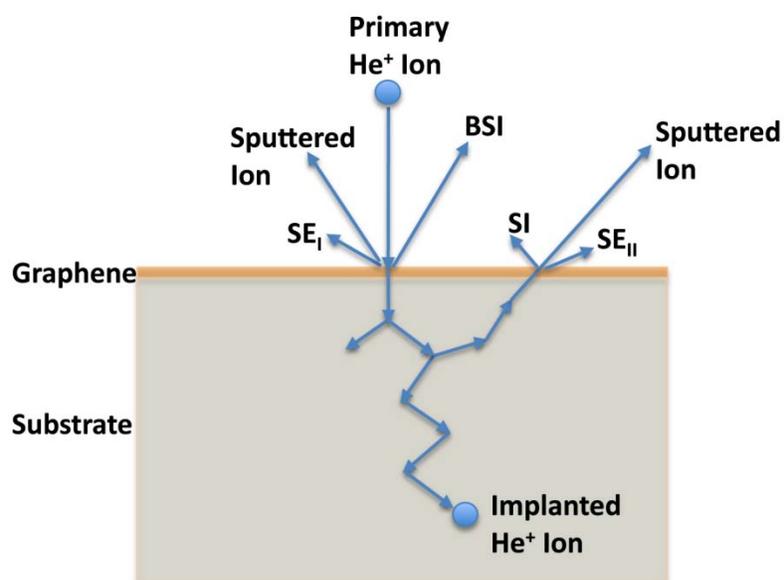

**Figure 2.** Schematic of the interactions of primary energetic He ions with a graphene layer on $SiO_2$ substrate, showing the production of secondary electrons ($SE_I$ at the primary beam and $SE_{II}$ at the secondary scattered ion exiting the surface), back scattered ions (BSI) and secondary ions (SI).

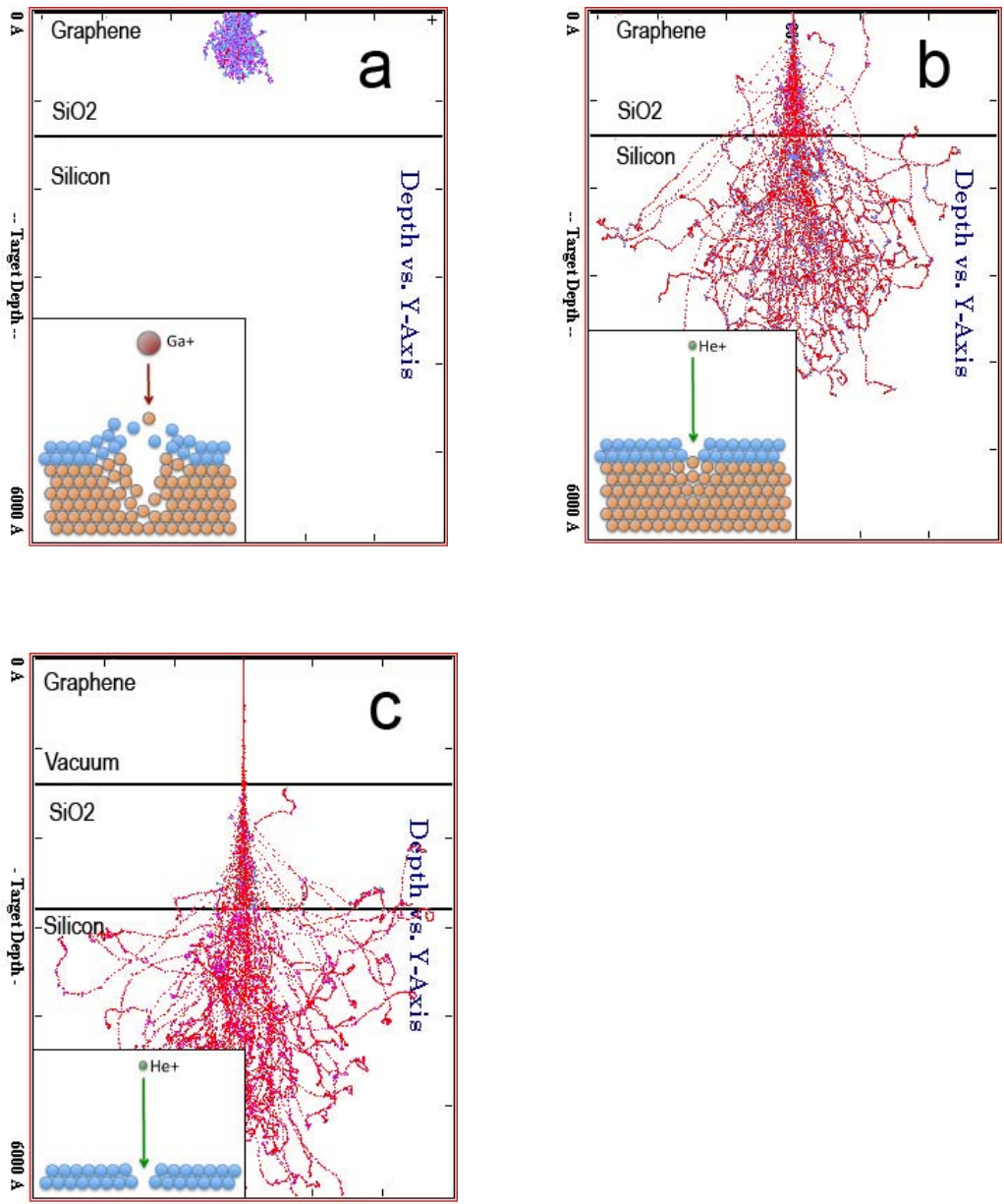

**Figure 3**. Simulations using TRIM software [10] comparing a) 30 kV Ga ions and b) 30kV He ions for range and trajectory in graphene layers on $SiO_2$ on silicon substrate. c) Range and trajectory of 30 kV He ions through a suspended graphene layer over vacuum, $SiO_2$ and silicon substrate. (Insets) Schematic comparison between Ga+ ion and He+ ion interaction with graphene samples from molecular dynamics simulations [11].

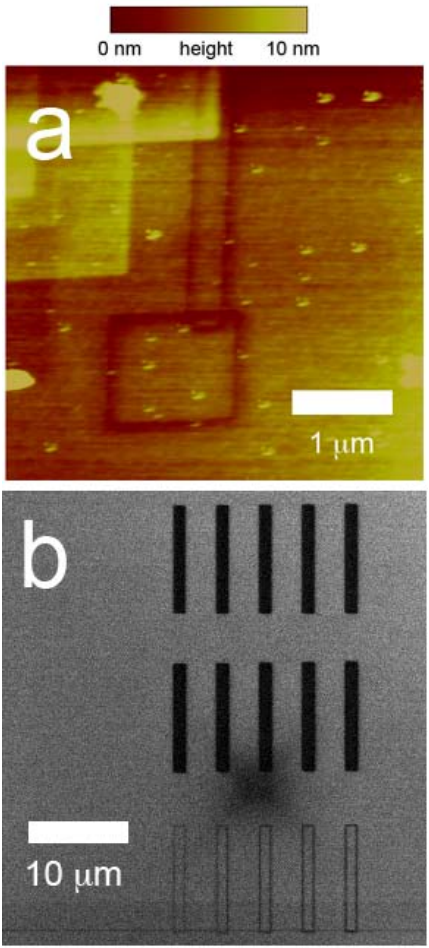

**Figure 4.** a) Test patterns written into a 285 nm $SiO_2$ film on silicon substrate as measured with AFM and b) Secondary electron image showing etching of boxes and line box patterns.

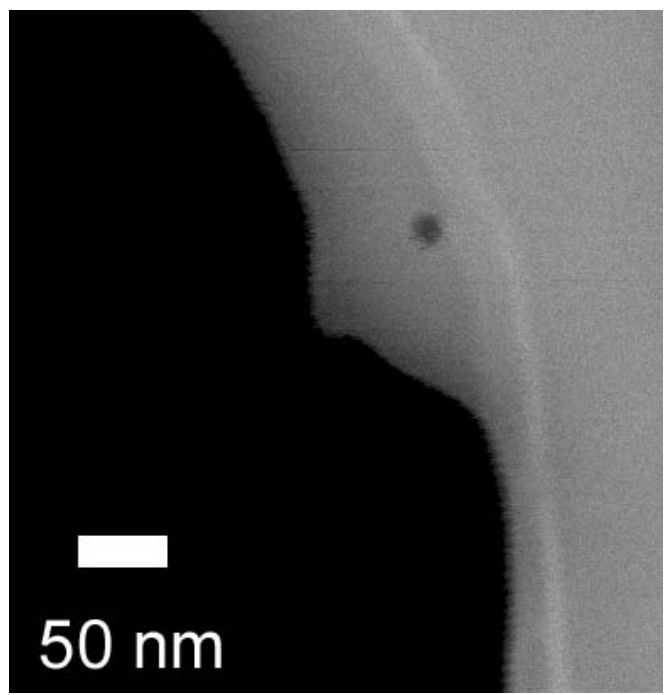

**Figure 5.** Helium ion micrograph imaged with the secondary electrons of a hole etched into a multi layer graphene flake (grey) on a SiO$_2$ substrate (black).

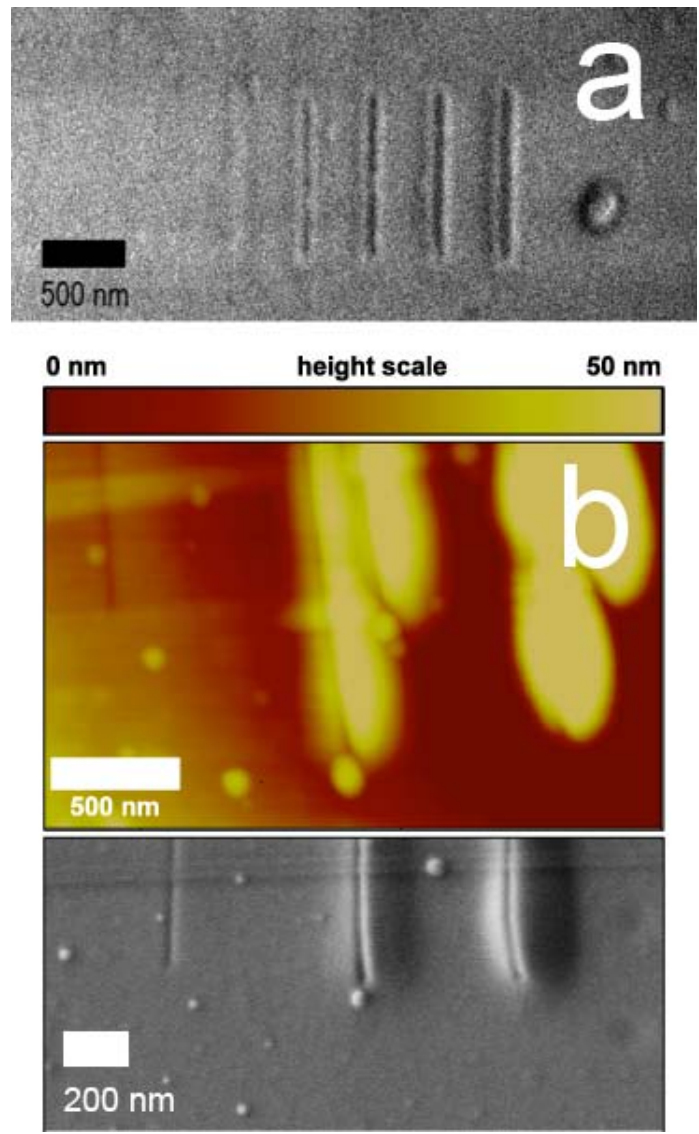

**Figure 6.** a) Helium ion micrographs of a dose variation test pattern in single-layer graphene, on SiO$_2$ substrate covering a range of linear doses from 3 nC/cm to 15 nC/cm in 3 nC/cm steps (left to right). b) AFM image with corresponding SEM image of a pattern etched with 20, 100 and 200 nC/cm line dose variation (from left to right).

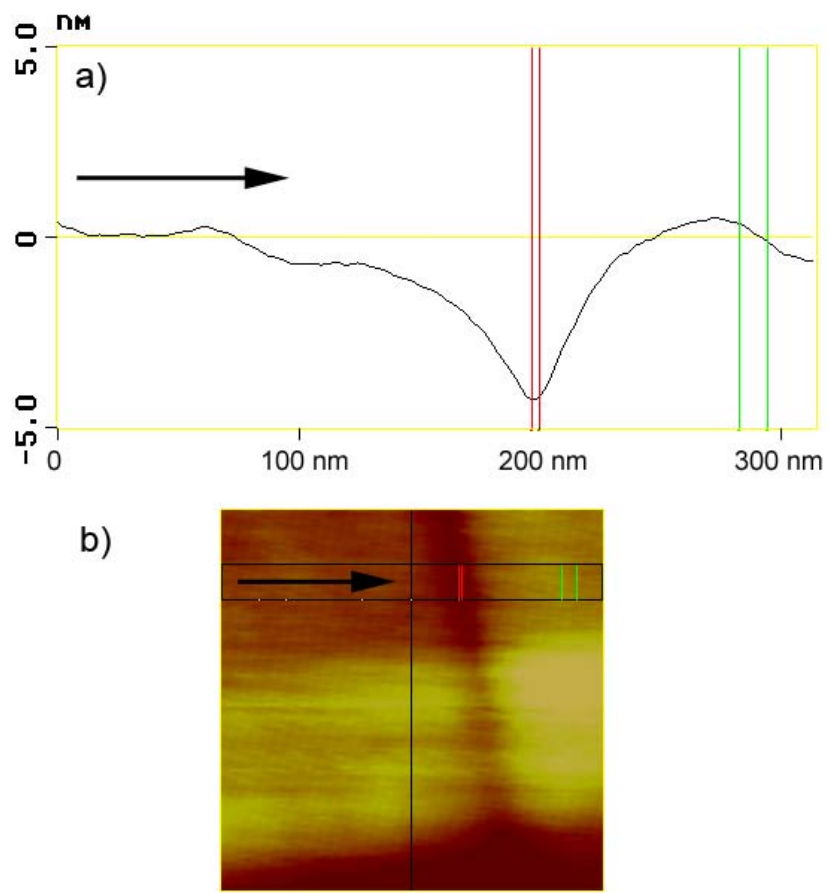

**Figure 7.** a) AFM step profile analysis for the single-layer graphene cut in Fig. 6(b) with a dose rate of 20 nC/cm resulting in a depth of 4 nm. b) AFM image used for the step profile. The profile was taken along the upper part of the image, in the position and direction indicated by the arrow.

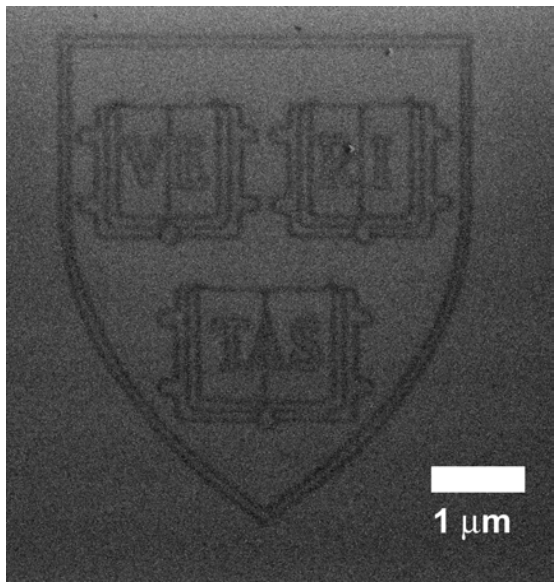

**Figure 8.** Helium ion micrograph of a high resolution Harvard University logo etched by HIL into multi-layer graphene flake.